\documentclass[
	aps, pra,
	superscriptaddress, twocolumn,
	floatfix,
	10pt
]{revtex4-1}

\usepackage{silence}
\usepackage[final]{graphicx}
\graphicspath{{./figures/}}
\usepackage{times,bbm,amsmath,amssymb}
\usepackage{microtype}
\usepackage{epsfig,color}
\usepackage[dvipsnames]{xcolor}
\usepackage{hyperref}
\hypersetup{
    colorlinks = true
}
\usepackage{cleveref}

\usepackage{libertine}

\usepackage{float,siunitx}
\usepackage[caption=false]{subfig}
\usepackage{booktabs}

\usepackage{multirow}
\usepackage{placeins}
\usepackage{relsize}  

\usepackage{gensymb}  
\usepackage[normalem]{ulem}

\usepackage{acronym}

\usepackage{easyReview}

\usepackage{tikz}
\usetikzlibrary{quantikz}
\usetikzlibrary{decorations.pathmorphing}
\usetikzlibrary{shapes.arrows}

\usepackage{standalone}

\usepackage{physics}

\newcommand{\bs}[1]{\boldsymbol{#1}}
\newcommand{\on}[1]{\operatorname{#1}}
\newcommand{\parTitle}[1]{\noindent{\color{Mahogany}(\emph{#1})}}

\newcommand{\CC}{\mathbb{C}}
\newcommand{\EE}{\mathbb{E}}
\newcommand{\NN}{\mathbb{N}}
\newcommand{\PP}{\mathbb{P}}
\newcommand{\RR}{\mathbb{R}}

\newcommand{\calC}{{\mathcal{C}}}

\newcommand{\calH}{{\mathcal{H}}}

\newcommand{\calS}{{\mathcal{S}}}

\newcommand{\calO}{{\mathcal{O}}}

\newcommand{\calW}{{\mathcal{W}}}

\newcommand{\HC}{\calH_\calC}
\newcommand{\HW}{\calH_\calW}

\renewcommand{\parTitle}[1]{}

\begin{document}
\title{Entanglement transfer, accumulation and retrieval \\ via quantum-walk-based qubit-qudit dynamics}

\author{Taira Giordani}
\affiliation{Dipartimento di Fisica, Sapienza Universit\`{a} di Roma, Piazzale Aldo Moro 5, I-00185 Roma, Italy}
\author{Luca  Innocenti} 
\affiliation{Department of Optics, Palack\'{y} University, 17. Listopadu 12, 771 46 Olomouc, Czech Republic}
\affiliation{Centre for Theoretical Atomic, Molecular, and Optical Physics,
School of Mathematics and Physics, Queen's University Belfast, BT7 1NN Belfast, United Kingdom}
\author{Alessia Suprano} 
\affiliation{Dipartimento di Fisica, Sapienza Universit\`{a} di Roma, Piazzale Aldo Moro 5, I-00185 Roma, Italy}
\author{Emanuele Polino} 
\affiliation{Dipartimento di Fisica, Sapienza Universit\`{a} di Roma, Piazzale Aldo Moro 5, I-00185 Roma, Italy}
\author{Mauro Paternostro} 
\affiliation{Centre for Theoretical Atomic, Molecular, and Optical Physics,
School of Mathematics and Physics, Queen's University Belfast, BT7 1NN Belfast, United Kingdom}
\author{Nicol\`o Spagnolo}
\affiliation{Dipartimento di Fisica, Sapienza Universit\`{a} di Roma, Piazzale Aldo Moro 5, I-00185 Roma, Italy}
\author{Fabio Sciarrino}
\affiliation{Dipartimento di Fisica, Sapienza Universit\`{a} di Roma, Piazzale Aldo Moro 5, I-00185 Roma, Italy}
\affiliation{Consiglio Nazionale delle Ricerche, Istituto dei sistemi Complessi (CNR-ISC), Via dei Taurini 19, 00185 Roma, Italy}
\author{Alessandro Ferraro}
\affiliation{Centre for Theoretical Atomic, Molecular, and Optical Physics,
School of Mathematics and Physics, Queen's University Belfast, BT7 1NN Belfast, United Kingdom}

\begin{abstract}
The generation and control of quantum correlations in high-dimensional systems is a major challenge in the present landscape of quantum technologies. Achieving such non-classical high-dimensional resources will potentially unlock enhanced capabilities for quantum cryptography, communication and computation. We propose a protocol that is able to attain entangled states of $d$-dimensional systems through a quantum-walk-based {\it transfer \& accumulate} mechanism involving coin and walker degrees of freedom. The choice of investigating quantum walks is motivated by their generality and versatility, complemented by their successful implementation in several physical systems. Hence, given the cross-cutting role of quantum walks across quantum information, our protocol potentially represents a versatile general tool to control high-dimensional entanglement generation in various experimental platforms. In particular, we illustrate a possible photonic implementation where the information is encoded in the orbital angular momentum and polarization degrees of freedom of single photons. 
\end{abstract}
\maketitle

\section{Introduction}

\parTitle{High-dimensional entanglement is relevant}
Quantum entanglement underpins many of the advantages promised by the technological advances in quantum information processors~\cite{horodecki2009quantum}.
Despite considerable research efforts have been devoted to achieving seamless generation and control of two-dimensional systems, it is known that two-dimensional entanglement entails limitations in a variety of settings~\cite{Greentree2004,Fujiwara2003,Kaszlikowski2000}.
When higher-dimensional entanglement is used --- for example in the context of quantum communication~\cite{cozzolino2019high} --- higher channel capacities can be achieved through superdense coding protocols \cite{liu2002general, grudka2002symmetric, hu2018beating}.  Quantum cryptography protocols enhanced by higher-dimensional entangled states achieve better performances in terms of key rates, noise resilience, and security~\cite{bechmannpasquinucci2000quantum, cerf2002security, Bruss2002, Karimipour2002, acin2003security, karimipour2002quantum,  durt2004security, groblacher2006experimental, huber2013weak, nunn2013largealphabet, mower2013highdimensional, lee2014entanglementbased, zhong2015photonefficient, Mirhosseini_2015}. Significant benefits can also be achieved in quantum error correction \cite{Chuang1997, Campbell2012, Duclos-Cianci2013, Michael2016} and fault-tolerant quantum computation \cite{bartlett2002quantum,ralph2007efficient, Lanyon2009, Campbell2014}.

\parTitle{HD entanglement is challenging, here is what we propose to do in a nutshell} The potential benefits of high-dimensional entanglement have stimulated a significant effort towards its generation, manipulation, and certification in various platforms including, in particular, optical systems~\cite{friis2019entanglement, erhard2020advances}. Despite significant experimental advances, the implementation of such tasks remains demanding, especially in light of the difficulties associated to controlling systems and  transformations in large Hilbert spaces.

In this paper, we show how to leverage controllable low-dimensional systems, together with special quantum devices acting as interfaces between systems of different dimensions, to realize an effective entanglement-transfer protocol from low- to high-dimensional degrees of freedom. Quantum correlations stored in two-dimensional degrees of freedom --- such as the polarizations of entangled photons --- can thus be passed into high-dimensional information carriers via suitable local interactions and measurements.

{We derive the general conditions under which such entanglement transfer is feasible.}
We then focus on the case of states producible by discrete-time one-dimensional quantum walks (QW)~\cite{aharonov1993quantum,nayak2000quantum,ambainis2001onedimensional,Kempe2003quantum,venegas-andraca2012quantum}. These model a natural type of interaction between hetero-dimensional systems, and are widely available in a variety of physical systems.
We study the conditions under which QW dynamics allow to transfer entanglement between coin and walker degrees of freedom, and prove the feasibility of accumulating entanglement in the high-dimensional system by repeatedly creating it and transfering it from the low-dimensional one.
This scheme constitutes a promising two-way interface to transfer reliably entanglement between different information carriers~\cite{Paternostro2004,Paternostro2004bis,Paternostro2004ter,Serafini2006,Adesso2010,Paternostro2009}.

A particularly suitable platform for the manipulation of high-dimensional systems, which has also been successful in demonstrating control of the QW dynamics, is embodied by the orbital angular momentum (OAM) of light. Recent experimental progress enabled by the growing capacity to prepare, manipulate and measure OAM states are opening up the possibility to explore the richness of high-dimensional Hilbert spaces for the sake of quantum information processing~\cite{erhard2018twisted}.
A protocol allowing to generate high-dimensional OAM states using a simple dynamics such as the one offered by QWs would therefore be a significant step forward towards the provision of \textit{on demand} high-dimensional entangled states.

\parTitle{Tentative outline}
The remainder of this paper is organized as follows. In~\cref{sec:overview} we overview the necessary background on QWs and OAM.
In~\cref{sec:entanglement_transfer_local_projections} we formalise the general conditions for the occurrence of entanglement transfer and study their solutions. We then specialise in~\cref{sec:entanglement_transfer_in_QWs} to the context set by QWs, and study -- in~\cref{sec:entanglement_accumulation} -- the possibility of accumulating entanglement in one degree of freedom by repeated applications of the entanglement-transfer protocol.
We conclude in~\cref{sec:experimental_proposal} by detailing a possible experimental implementation of the protocol in the framework of OAM-based implementation of the QW dynamics.

\section{Background}
\label{sec:overview}

\parTitle{Background on QWs}
Discrete-time QWs embody a widely studied type of interaction between a two-dimensional ``\textit{coin}'' degree of freedom, and a high-dimensional ``\textit{walker}'' one
~\cite{aharonov1993quantum,nayak2000quantum,ambainis2001onedimensional,Kempe2003quantum,venegas-andraca2012quantum}.
Despite their simplicity, QWs allow to engineer effectively a broad range of evolutions \cite{kitagawa2012observation, sansoni2012twoparticle, cardano_zak_2017, banchi2016quench}.
Recently, some of us demonstrated the potential of a QW-based architecture to flexibly implement quantum state engineering of a single OAM~\cite{innocenti2017quantum,giordani2018experimental}, as well as the machine-learning-enhanced classification of hybrid polarization-OAM states of light~\cite{giordani2020machine}.
A possible physical embodiment of such QW dynamics uses polarization and OAM of single photons, playing the roles of the coin and the walker degrees of freedom, respectively, with waveplates to implement the coin operations and q-plates~\cite{marrucci2006optical} to implement the controlled-shift.
State engineering protocols leveraging QWs in this setting were previously designed and demonstrated in Refs.~\cite{innocenti2017quantum,giordani2019experimental,giordani2020machine}.

More precisely, QWs are defined in a bipartite coin-walker space $\HC\otimes\HW$, where ${\cal H}_{{\cal C}({\cal W})}$ denotes the coin (walker) space.
We assume $\dim(\HC)=2$.
The evolution is defined by the repeated action of a unitary \textit{walk operation} $\calW_\calC\equiv \calS(\calC\otimes I)$, which comprises the sequential action of a \textit{controlled-shift} operation $\calS$, and a \textit{coin flipping} operation $\calC$.
The coin flipping operation acts locally on the coin space, while the controlled-shift changes the state of the walker conditionally to the state of the coin:
\begin{equation}
    \calS \equiv \sum_k (
        \PP_\uparrow\otimes \ketbra{k}{k} +
        \PP_\downarrow\otimes \ketbra{k+1}{k}
    ),
\end{equation}
where $\{\ket\uparrow,\ket\downarrow\}$ form a basis for $\HC$, $\{\ket k\}_{k\ge0}$ spans $\HW$, and we introduced the notation $\PP_\psi\equiv\ketbra\psi$.

\parTitle{Pairs of QWs}
The state space we are interested in consists of two  pairs of QWs, so that the overall system of coins and walkers lives in the four-partite space $\calH\equiv \calH^{(1)}\otimes\calH^{(2)}$, with
$\calH^{(i)}\equiv \HC^{(i)}\otimes\HW^{(i)}$,
and $\HC^{(i)}, \HW^{(i)}$ accommodating coin and walker of the $i^{\text th}$ party, respectively ($i=1,2$).
Given $\ket\Psi\in\calH$, we apply $\calW_\calC$ locally on $\calH^{(1)}$ and $\calH^{(2)}$.
This, in general, entangles each coin with the respective walker \cite{Viera2013, Gratsea_2020}. 
In the next sections, we will describe how to use this QW dynamics to transfer entanglement from the two-coin subspace to the two-walker one, using only local operations on the coins.
In an optical setup, this process will transfer the initial entanglement encoded in a polarization state to the two OAM degrees of freedom. The process can be iterated to transfer more entanglement from the polarizations to the OAMs.

\begin{figure*}[ht]
    \centering 
    \includegraphics[width=0.9\textwidth]{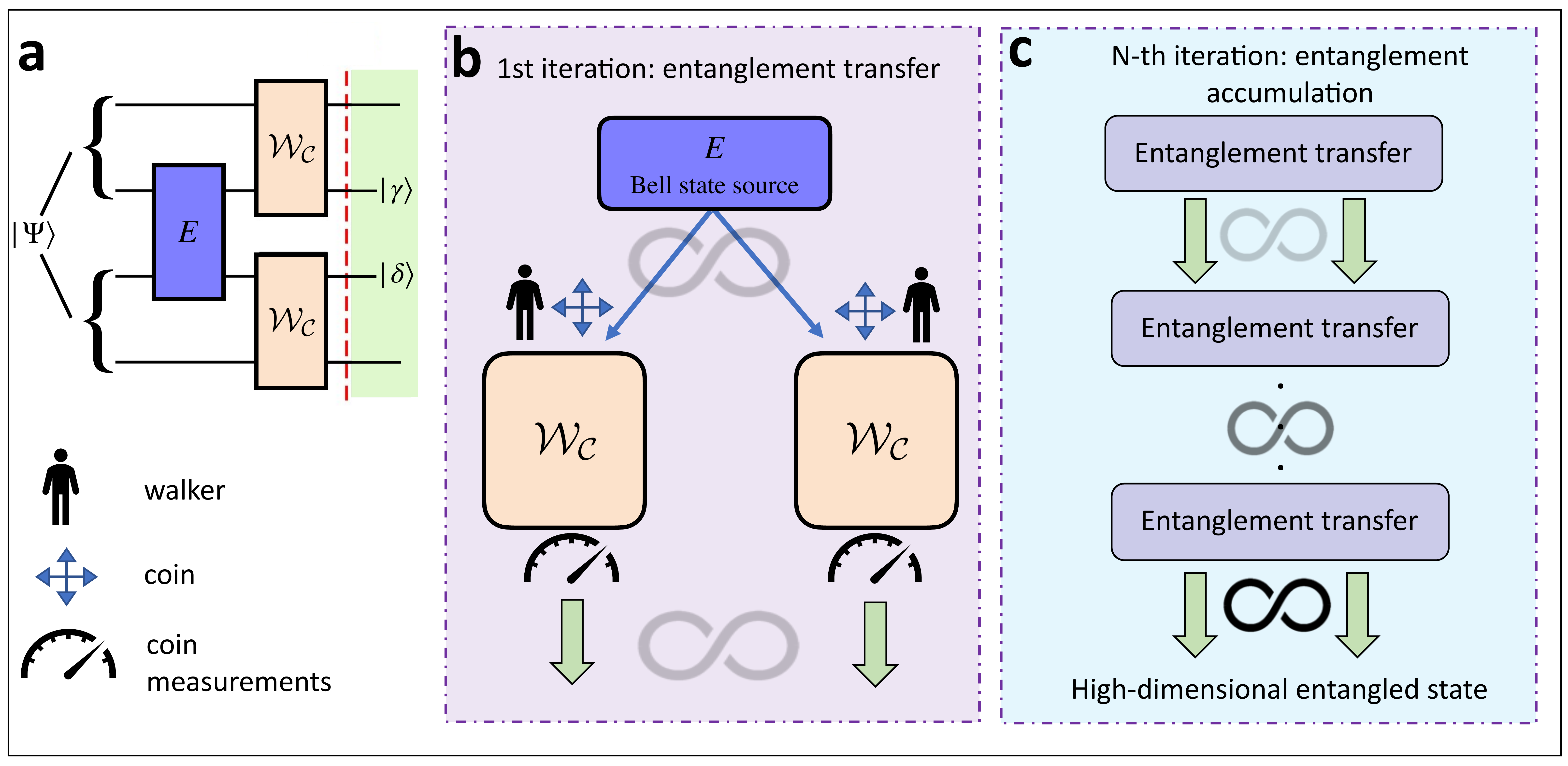}
    \caption{
        {\bf a} Entanglement transfer unit. The system is composed by two particles, 1 and 2, equipped with a $\ell$-dimensional degree of freedom, which will be instrumental to the protocol, and an additional $d$-dimensional degree of freedom. The entanglement transfer protocol requires a first operation $E$ that generates entanglement between the $\ell$-dimensional subsystems. Then we have two {\it local} operations -- with respect to the 1-vs-2 bipartition -- that correlates the inner degrees of freedoms of each particle and realizes the $\ell$-$d$ dynamics. In the end, local measurements allow to transfer the entanglement stored in the initial state to the reduced state of the $d$-dimensional sub-systems. We consider explicitly the case of $\ell=2$ (qubits) and local operations embodied by the walk operations ${\cal W}_{\cal C}$. Indeed, a discrete-time quantum walks framework 
        offers a very natural encoding of this dynamics: in such embodiment, the  coin particle would 
        codify the $\ell=2$-dimensional system, with the qudit being provided by the position degrees of freedom of the walker. Assuming initially maximally entangled states of the qubits, a single iteration of our protocol would be able to transfer one ebit of entanglement at most. By repeating the use of this unit, high-dimensional entangled states can be generated in the $d$-dimensional degrees of freedom. Furthermore the entanglement stored in such degrees of freedom can be retrieved by same operations and transferred back to the two-qubit state. 
         {\bf b} Conceptual scheme for the transfer from a Bell state in the coin degree of freedom to the two walkers position space after quantum walks and local coin measurements. {\bf c}  Protocol iteration and entanglement accumulation in the high-dimensional space of the two quantum walkers.}
    \label{fig:conceptual_scheme}
\end{figure*}

\section{Entanglement transfer via local projections}
\label{sec:entanglement_transfer_local_projections}

\parTitle{Problem statement}
We now address the challenge of \textit{transferring} entanglement across different degrees of freedom using solely local projections. More precisely, we consider \textit{four-partite} states $\ket\Psi\in\calH$, and ask when, via local projections, it is possible to transfer, or ``\textit{focus}'', the entanglement into the bipartition $\HW^{(1)}\otimes\HW^{(2)}$.
We thus look for conditions ensuring the existence of states $\ket{\gamma}\in\calH^{(1)}_{\calC}$ and $\ket\delta\in\calH^{(2)}_{\calC}$ such that the entanglement of $\ket\Psi$ in the bipartition $\calH^{(1)}\otimes\calH^{(2)}$ is preserved in the projected state $\braket*{\gamma,\delta}{\Psi}\in\calH^{(1)}_{\calW}\otimes\calH^{(2)}_{\calW}$.
A schematic description of this formal scenario is given in Fig.~\ref{fig:conceptual_scheme}a.
Note that such entanglement transfer is not always possible. 
It is therefore pivotal to find the conditions making such protocol viable.
{It is worth noting that, when probabilistic operations are allowed (as in the case of projections), even restricting to {local} operations, the amount of entanglement \textit{can} be increased~\cite{nielsen2001majorization,nielsen1999conditions,vidal1999entanglement}. Such process of effective \textit{entanglement distillation} comes, however, at the expense of lowered success probabilities. We focus here on the case where we want to preserve, not enhance, the entanglement in a given state. In this case, it is also possible to achieve entanglement transfer deterministically, when there is a complete basis of projections each element of which achieves entanglement transfer.}

\parTitle{We are actually dealing with two problems at once}
We can break down the task at hand into two independent sub-problems, which we will refer to as \textit{transferability conditions}: on the one hand, transferring the entanglement from $\calH^{(1)}\otimes\calH^{(2)}$ to $\HW^{(1)}\otimes \calH^{(2)}$, and on the other hand, transferring the entanglement from $\HW^{(1)}\otimes \calH^{(2)}$ to $\HW^{(1)}\otimes\HW^{(2)}$.
The achievability of these two tasks will be referred to with $\text{TC}_1$ and $\text{TC}_2$, respectively.
It is worth stressing that, while throughout the paper we will always focus our discussion on $\text{TC}_1$, all results hold analogously for $\text{TC}_2$ when instead of projecting in the space $\HC^{(1)}$ we project in $\HC^{(2)}$.

\parTitle{Conditions for entanglement transfer}
To frame the problem more precisely,
consider a state $\ket\Psi\in\calH$ with Schmidt decomposition
\begin{equation}
	\ket\Psi= \sum_k \sqrt{p_k} \ket{u_k}\ket{v_k},
\end{equation}
where $\sum_k p_k=1$, $\ket{u_k}\in{\cal H}^{(1)}$ and $\ket{v_k}\in{\cal H}^{(2)}$.
To achieve $\text{TC}_1$ we want a state $\ket\gamma\in\HC^{(1)}$ such that the corresponding projected state $\ket{\Psi_\gamma}\in\HW^{(1)}\otimes\calH^{(2)}$ contains the same amount of entanglement, in the bipartition $\HW^{(1)}\otimes\calH^{(2)}$, as that initially in $\ket\Psi$.
In general, we have
\begin{equation}
    \ket{\Psi_\gamma} = \frac{1}{\sqrt{p_{\rm proj}}}
    \sum_k \sqrt{p_k q_k} \ket{\tilde u_k}\ket{v_k},
    \label{eq:postproj_state}
\end{equation}
where
$\sqrt{q_k}\ket{\tilde u_k}= \braket\gamma{u_k}\in{\cal H}^{(1)}$
and
$p_{\rm proj} = \sum_k p_k q_k$.
We distinguish between three different scenarios:
\begin{enumerate}
\item[(1)]If the states $\ket{\tilde u_k}$ are not orthogonal, then some information about which $k$ the state is in leaks through the coin projection, and some  entanglement is thus degraded. This will be shown formally in Appendix~\ref{appA}.\\
\item[(2)] If the states $\ket{\tilde u_k}$ are orthogonal, but the corresponding projection probabilities $q_k$ are uneven, then again the entanglement in $\ket{\Psi_\gamma}$ is smaller than that in $\ket\Psi$.\\
\item[(3)] If the states $\ket{\tilde u_k}$ are orthogonal, \emph{and} $q_k=p_{\on{proj}}$ for all $k$, \textit{then} projecting onto $\ket\gamma$ fully preserves the initial entanglement.
\end{enumerate}
Note that situation (3) is thus a necessary and sufficient condition for entanglement transferability without degradation,
as if $\braket{\tilde u_j}{\tilde u_k}=\delta_{jk}$ and $q_k=p_{\on{proj}}$ then~ Eq. \eqref{eq:postproj_state} is the Schmidt decomposition of $\ket{\Psi_\gamma}$, and therefore the Schmidt coefficients of $\ket{\Psi_\gamma}$ are (in the relevant bipartition) the same as those of $\ket{\Psi}$.
On the other hand, if (3) is not satisfied, then the projection results in the degradation of some of the entanglement, as shown in Appendix A.

\begin{figure}[b]
    \centering
    \tikzset{every picture/.style={line width=0.75pt}} 

\begin{tikzpicture}[x=0.75pt,y=0.75pt,yscale=-1,xscale=1]

\draw [draw=red, decorate, decoration=snake] (186,62) -- (186,135) ;
\draw  [draw=red, decorate, decoration=snake]  (256,60) -- (256,135) ;
\draw [draw=blue, decorate, decoration=snake] (202.54,100.92) -- (238.66,100.44) ;
\draw  [fill={rgb, 255:red, 0; green, 0; blue, 0 }  ,fill opacity=1 ] (181,65) .. controls (181,62.24) and (183.24,60) .. (186,60) .. controls (188.76,60) and (191,62.24) .. (191,65) .. controls (191,67.76) and (188.76,70) .. (186,70) .. controls (183.24,70) and (181,67.76) .. (181,65) -- cycle ;
\draw  [fill={rgb, 255:red, 0; green, 0; blue, 0 }  ,fill opacity=1 ] (181,135) .. controls (181,132.24) and (183.24,130) .. (186,130) .. controls (188.76,130) and (191,132.24) .. (191,135) .. controls (191,137.76) and (188.76,140) .. (186,140) .. controls (183.24,140) and (181,137.76) .. (181,135) -- cycle ;
\draw  [fill={rgb, 255:red, 0; green, 0; blue, 0 }  ,fill opacity=1 ] (251,65) .. controls (251,62.24) and (253.24,60) .. (256,60) .. controls (258.76,60) and (261,62.24) .. (261,65) .. controls (261,67.76) and (258.76,70) .. (256,70) .. controls (253.24,70) and (251,67.76) .. (251,65) -- cycle ;
\draw  [fill={rgb, 255:red, 0; green, 0; blue, 0 }  ,fill opacity=1 ] (251,135) .. controls (251,132.24) and (253.24,130) .. (256,130) .. controls (258.76,130) and (261,132.24) .. (261,135) .. controls (261,137.76) and (258.76,140) .. (256,140) .. controls (253.24,140) and (251,137.76) .. (251,135) -- cycle ;
\draw  [color={rgb, 255:red, 139; green, 87; blue, 42 }  ,draw opacity=1 ][line width=1.5]  (177,126) -- (195,126) -- (195,144) -- (177,144) -- cycle ;
\draw [color={rgb, 255:red, 139; green, 87; blue, 42 }  ,draw opacity=1 ][line width=1.5]    (185.8,143.8) -- (185.82,151.33) ;
\draw [shift={(185.83,155.33)}, rotate = 269.83] [fill={rgb, 255:red, 139; green, 87; blue, 42 }  ,fill opacity=1 ][line width=0.08]  [draw opacity=0] (6.97,-3.35) -- (0,0) -- (6.97,3.35) -- cycle    ;
\draw   (300.67,97.17) -- (320.57,97.17) -- (320.57,94.67) -- (333.83,99.67) -- (320.57,104.67) -- (320.57,102.17) -- (300.67,102.17) -- cycle ;
\draw [draw=blue, decorate, decoration=snake] (400.4,100) -- (353.8,100) ;
\draw [draw=red, decorate, decoration=snake] (417.8,60) -- (417.8,135) ;
\draw  [fill={rgb, 255:red, 0; green, 0; blue, 0 }  ,fill opacity=1 ] (348.8,100) .. controls (348.8,97.24) and (351.04,95) .. (353.8,95) .. controls (356.56,95) and (358.8,97.24) .. (358.8,100) .. controls (358.8,102.76) and (356.56,105) .. (353.8,105) .. controls (351.04,105) and (348.8,102.76) .. (348.8,100) -- cycle ;
\draw  [fill={rgb, 255:red, 0; green, 0; blue, 0 }  ,fill opacity=1 ] (412.8,65) .. controls (412.8,62.24) and (415.04,60) .. (417.8,60) .. controls (420.56,60) and (422.8,62.24) .. (422.8,65) .. controls (422.8,67.76) and (420.56,70) .. (417.8,70) .. controls (415.04,70) and (412.8,67.76) .. (412.8,65) -- cycle ;
\draw  [fill={rgb, 255:red, 0; green, 0; blue, 0 }  ,fill opacity=1 ] (412.8,135) .. controls (412.8,132.24) and (415.04,130) .. (417.8,130) .. controls (420.56,130) and (422.8,132.24) .. (422.8,135) .. controls (422.8,137.76) and (420.56,140) .. (417.8,140) .. controls (415.04,140) and (412.8,137.76) .. (412.8,135) -- cycle ;
\draw [color=red] (401.22,100.41) .. controls (401.58,74.76) and (409.29,54.07) .. (418.42,54.2) .. controls (427.55,54.33) and (434.66,75.23) .. (434.29,100.88) .. controls (433.93,126.53) and (426.22,147.22) .. (417.09,147.09) .. controls (407.96,146.96) and (400.85,126.06) .. (401.22,100.41) -- cycle ;
\draw [color=red] (169.46,100.44) .. controls (169.83,74.79) and (177.53,54.11) .. (186.66,54.24) .. controls (195.8,54.37) and (202.9,75.27) .. (202.54,100.92) .. controls (202.17,126.57) and (194.47,147.26) .. (185.34,147.12) .. controls (176.2,146.99) and (169.1,126.09) .. (169.46,100.44) -- cycle ;
\draw [color=red] (238.66,100.44) .. controls (239.03,74.79) and (246.73,54.11) .. (255.86,54.24) .. controls (265,54.37) and (272.1,75.27) .. (271.74,100.92) .. controls (271.37,126.57) and (263.67,147.26) .. (254.54,147.12) .. controls (245.4,146.99) and (238.3,126.09) .. (238.66,100.44) -- cycle ;

\draw (136.8,47.4) node [anchor=north west][inner sep=0.75pt]    {$\mathcal{H}^{( 1)}_{\mathcal{W}}$};
\draw (267,47.4) node [anchor=north west][inner sep=0.75pt]    {$\mathcal{H}^{( 2)}_{\mathcal{W}}$};
\draw (136.8,125.4) node [anchor=north west][inner sep=0.75pt]    {$\mathcal{H}^{( 1)}_{\mathcal{C}}$};
\draw (267,125.4) node [anchor=north west][inner sep=0.75pt]    {$\mathcal{H}^{( 2)}_{\mathcal{C}}$};
\draw (174.13,156.73) node [anchor=north west][inner sep=0.75pt]    {$\langle \gamma |$};
\draw (338.8,71.4) node [anchor=north west][inner sep=0.75pt]    {$\mathcal{H}^{( 1)}_{\mathcal{W}}$};
\draw (430.8,47.4) node [anchor=north west][inner sep=0.75pt]    {$\mathcal{H}^{( 2)}_{\mathcal{W}}$};
\draw (430.8,125.4) node [anchor=north west][inner sep=0.75pt]    {$\mathcal{H}^{( 2)}_{\mathcal{C}}$};

\end{tikzpicture}
    \caption{Pictorial representation of the first transferability procedure.
    Given a state which is entangled with respect to the bipartition $\calH^{(1)}\otimes\calH^{(2)}$, we apply a local projection $\ket\gamma$ which preserves the entanglement between the two spaces.
    Condition~\eqref{eq:transferability_condition} determines when such a projection exists.
    }
    \label{fig:TC1_general_condition_scheme}
\end{figure}
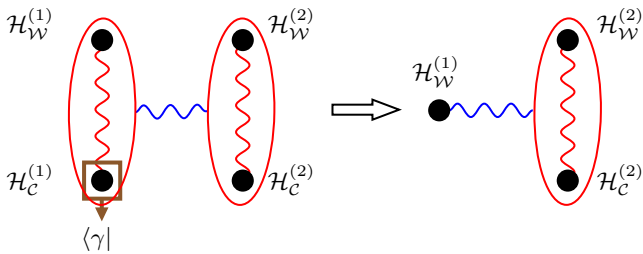

\parTitle{Summary of conclusions}
Therefore, we achieve transferability if $\ket\gamma$ is such that
$\braket{\gamma}{u_k}/{\sqrt{p_{\on{proj}}}}$ are orthonormal vectors.
An equivalent --- if less explicit --- condition for transferability is the requirement
\begin{equation}
    \tilde\sigma(\tr_2(\PP_{\Psi_\gamma})) =
    \tilde\sigma\left(
        \tr_2(\PP_\Psi)
    \right),
    \label{eq:transferability_condition}
\end{equation}
where $\tilde\sigma(A)\equiv\sigma(A)\setminus\{0\}$ and $\sigma(A)$ is the set of eigenvalues of $A$.
This is a \emph{necessary and sufficient} condition for transferability, as Eq.~\eqref{eq:transferability_condition} is equivalent to requesting that the Schmidt coefficients of $\ket{\Psi_\gamma}$ are the same as those of $\ket{\Psi}$.
In Fig.~\ref{fig:TC1_general_condition_scheme} we present a pictorial description of what $\on{TC}_1$ allows to achieve.
It is worth noting that, while Eq.~\eqref{eq:transferability_condition} is required to fully transfer entanglement, it is still possible to transfer \textit{some} degree of entanglement if the vectors $\braket{\gamma}{u_k}$ are not fully orthogonal, or the projection probabilities are unequal.

\parTitle{Relations with entanglement swapping}
This problem can be understood as a more restrictive version of entanglement swapping.
Such protocol~\cite{zukowski1993eventreadydetectors} deals with a four-partite system in the Hilbert space $\otimes_{j}\calH_{j}~(j=A,B,C,D)$, whose state is  separable in the bipartition $(AB)$-vs-$(CD)$ but entangled in the subsystems $A-B$ {\it and} $C-D$. The goal of entanglement swapping is to achieve  entanglement in the state of the $A-D$ compound by performing projective measurements on $B-C$. This is possible for instance by implementing a Bell measurement over the joint state of $B$ and $C$. Clearly, the problem is analogous to ours, except that we only allow \emph{local} operations on $B$ and $C$. Notably, the use of a Bell measurement is not available in our setting.

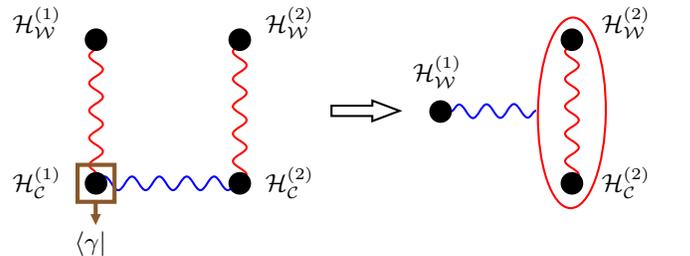
\begin{figure}[b]
    \centering
    \tikzset{every picture/.style={line width=0.75pt}} 

\begin{tikzpicture}[x=0.75pt,y=0.75pt,yscale=-1,xscale=1]
\draw  [draw=red, decorate, decoration=snake]  (45,62) -- (45,135) ;
\draw  [draw=red, decorate, decoration=snake]  (115,60) -- (115,135) ;
\draw  [draw=blue, decorate, decoration=snake]  (45,135) -- (115,135) ;
\draw  [fill={rgb, 255:red, 0; green, 0; blue, 0 }  ,fill opacity=1 ] (40,65) .. controls (40,62.24) and (42.24,60) .. (45,60) .. controls (47.76,60) and (50,62.24) .. (50,65) .. controls (50,67.76) and (47.76,70) .. (45,70) .. controls (42.24,70) and (40,67.76) .. (40,65) -- cycle ;
\draw  [fill={rgb, 255:red, 0; green, 0; blue, 0 }  ,fill opacity=1 ] (40,135) .. controls (40,132.24) and (42.24,130) .. (45,130) .. controls (47.76,130) and (50,132.24) .. (50,135) .. controls (50,137.76) and (47.76,140) .. (45,140) .. controls (42.24,140) and (40,137.76) .. (40,135) -- cycle ;
\draw  [fill={rgb, 255:red, 0; green, 0; blue, 0 }  ,fill opacity=1 ] (110,65) .. controls (110,62.24) and (112.24,60) .. (115,60) .. controls (117.76,60) and (120,62.24) .. (120,65) .. controls (120,67.76) and (117.76,70) .. (115,70) .. controls (112.24,70) and (110,67.76) .. (110,65) -- cycle ;
\draw  [fill={rgb, 255:red, 0; green, 0; blue, 0 }  ,fill opacity=1 ] (110,135) .. controls (110,132.24) and (112.24,130) .. (115,130) .. controls (117.76,130) and (120,132.24) .. (120,135) .. controls (120,137.76) and (117.76,140) .. (115,140) .. controls (112.24,140) and (110,137.76) .. (110,135) -- cycle ;
\draw   (159.67,97.17) -- (179.57,97.17) -- (179.57,94.67) -- (192.83,99.67) -- (179.57,104.67) -- (179.57,102.17) -- (159.67,102.17) -- cycle ;
\draw  [color={rgb, 255:red, 139; green, 87; blue, 42 }  ,draw opacity=1 ][line width=1.5]  (36,126) -- (54,126) -- (54,144) -- (36,144) -- cycle ;
\draw [color={rgb, 255:red, 139; green, 87; blue, 42 }  ,draw opacity=1 ][line width=1.5] (44.8,143.8) -- (44.82,152.33) ;
\draw [shift={(44.83,155.33)}, rotate = 269.83] [fill={rgb, 255:red, 139; green, 87; blue, 42 }  ,fill opacity=1 ][line width=0.08]  [draw opacity=0] (5.36,-2.57) -- (0,0) -- (5.36,2.57) -- cycle    ;
\draw [draw=blue, decorate, decoration=snake] (259.4,99.6) -- (212.8,100) ;
\draw  [draw=red, decorate, decoration=snake]  (276.8,60) -- (276.8,135) ;
\draw  [fill={rgb, 255:red, 0; green, 0; blue, 0 }  ,fill opacity=1 ] (207.8,100) .. controls (207.8,97.24) and (210.04,95) .. (212.8,95) .. controls (215.56,95) and (217.8,97.24) .. (217.8,100) .. controls (217.8,102.76) and (215.56,105) .. (212.8,105) .. controls (210.04,105) and (207.8,102.76) .. (207.8,100) -- cycle ;
\draw  [fill={rgb, 255:red, 0; green, 0; blue, 0 }  ,fill opacity=1 ] (271.8,65) .. controls (271.8,62.24) and (274.04,60) .. (276.8,60) .. controls (279.56,60) and (281.8,62.24) .. (281.8,65) .. controls (281.8,67.76) and (279.56,70) .. (276.8,70) .. controls (274.04,70) and (271.8,67.76) .. (271.8,65) -- cycle ;
\draw  [fill={rgb, 255:red, 0; green, 0; blue, 0 }  ,fill opacity=1 ] (271.8,135) .. controls (271.8,132.24) and (274.04,130) .. (276.8,130) .. controls (279.56,130) and (281.8,132.24) .. (281.8,135) .. controls (281.8,137.76) and (279.56,140) .. (276.8,140) .. controls (274.04,140) and (271.8,137.76) .. (271.8,135) -- cycle ;
\draw [draw=red]  (260.22,100.41) .. controls (260.58,74.76) and (268.29,54.07) .. (277.42,54.2) .. controls (286.55,54.33) and (293.66,75.23) .. (293.29,100.88) .. controls (292.93,126.53) and (285.22,147.22) .. (276.09,147.09) .. controls (266.96,146.96) and (259.85,126.06) .. (260.22,100.41) -- cycle ;

\draw (2.8,47.4) node [anchor=north west][inner sep=0.75pt]    {$\mathcal{H}^{( 1)}_{\mathcal{W}}$};
\draw (126,47.4) node [anchor=north west][inner sep=0.75pt]    {$\mathcal{H}^{( 2)}_{\mathcal{W}}$};
\draw (2.8,125.4) node [anchor=north west][inner sep=0.75pt]    {$\mathcal{H}^{( 1)}_{\mathcal{C}}$};
\draw (126,125.4) node [anchor=north west][inner sep=0.75pt]    {$\mathcal{H}^{( 2)}_{\mathcal{C}}$};
\draw (33.13,156.73) node [anchor=north west][inner sep=0.75pt]    {$\langle \gamma |$};
\draw (197.8,71.4) node [anchor=north west][inner sep=0.75pt]    {$\mathcal{H}^{( 1)}_{\mathcal{W}}$};
\draw (289.8,47.4) node [anchor=north west][inner sep=0.75pt]    {$\mathcal{H}^{( 2)}_{\mathcal{W}}$};
\draw (289.8,125.4) node [anchor=north west][inner sep=0.75pt]    {$\mathcal{H}^{( 2)}_{\mathcal{C}}$};
\end{tikzpicture}
    \caption{
        Like Fig.~\ref{fig:TC1_general_condition_scheme}, but for states in which the entanglement is only due to pre-shared entanglement between the coins. These are the types of states at the first entanglement accumulation step.
    }
    \label{fig:TC1_condition_scheme}
\end{figure}

\section{Entanglement transfer through Quantum-Walk dynamics}
\label{sec:entanglement_transfer_in_QWs}

In~\cref{sec:entanglement_transfer_local_projections} we discussed the general problem of transferring entanglement by means of local projections.
Most notably we made no assumption on the inner structure of correlations in $\calH^{(i)}$, nor we specified the dimensionality of the entanglement in the bipartition $\calH^{(1)}\otimes\calH^{(2)}$. The framework and results set up so far thus also apply to 
cases where some pre-existing entanglement exists between the walkers' degrees of freedom.
We now specialize to the case $\dim\HC^{(i)}=2$, which applies directly to QWs with two-dimensional coins.
More precisely, in ~\cref{subsec:generalsolution_firstiteration_2Dcoin} we consider states in which $\calH^{(1)}$ and $\calH^{(2)}$ are only entangled through their coin spaces (as in ~\cref{fig:TC1_condition_scheme}).
In~\cref{subsec:analytical_results_QWs} we then apply these results to the output states obtained from QW dynamics.

\subsection{Entanglement transfer via two-dimensional coins}
\label{subsec:generalsolution_firstiteration_2Dcoin}

\parTitle{Problem setting}
Consider a state $\ket\Psi\in\calH^{(1)}\otimes\calH^{(2)}$ which is entangled only via its coin spaces (or more generally, a state having rank $2$), as in Fig~\ref{fig:TC1_condition_scheme}. The corresponding reduced state reads
\begin{equation}
	\rho= p_1 \PP_u + p_2 \PP_v, \,\,p_1+p_2=1,
	\label{eq:reduced_state_twodimcase}
\end{equation}
for a pair of orthonormal states $\{\ket u,\ket v\}\in\calH^{(1)}$.
As discussed in~\cref{sec:entanglement_transfer_local_projections}, to achieve maximal entanglement transfer we need a projection onto a state $\ket\gamma$ satisfying $\on{TC}_1$, \emph{i.e.} fulfilling Eq.~\eqref{eq:transferability_condition}.
This is equivalent to requiring $\braket{\tilde u}{\tilde v}=0$ where $\braket{\gamma}{j}=\sqrt{p_{\on{proj}}}\ket{\tilde j}~(j=u,v)$.
Explicitly, these amount to the conditions
\begin{equation}
	\mel{\gamma}{\tr_{\calW}(\ketbra{u}{v})}{\gamma} = 0,
\end{equation}
and
$\mel{\gamma}{\tr_{\calW}(\ketbra{u})}{\gamma} =
\mel{\gamma}{\tr_{\calW}(\ketbra{v})}{\gamma} = p_{\on{proj}}$.
We show in Appendix~\ref{appB} that it is always possible to find a state $\ket\gamma$ that preserves the orthogonality.
To satisfy condition $\on{TC}_1$, one then only has to verify that the projection probabilities are equal.

\subsection{Entanglement transfer with coined QWs}
\label{subsec:analytical_results_QWs}

\parTitle{Section outline}
We now apply the results of the previous section to the specific quantum states resulting from coined QWs.
As in~\cref{subsec:generalsolution_firstiteration_2Dcoin}, we first 
assume that the overall state is entangled with respect to the bipartition $\calH^{(1)}\otimes\calH^{(2)}$ only via its coin spaces (see Fig.~\ref{fig:TC1_condition_scheme}). 
We thus take the initial full state of the form
\begin{equation}
\ket{\Psi}=    \sqrt{p_1} \ket{\uparrow,1}\otimes\ket{\uparrow,1} +
    \sqrt{p_2} \ket{\downarrow,1}\otimes\ket{\downarrow,1},
\end{equation}
for some coefficients $p_1,p_2\ge0$ with $p_1+p_2=1$.
Focusing on $\calH^{(1)}$, we thus see that the initial states upon which the QW operates are $\ket{\uparrow,1}$ and $\ket{\downarrow,1}$.

\parTitle{Single step}
A single QW step with coin operation  ${\cal C}$ 
amounts to the evolution
\begin{equation}
\begin{aligned}
	\ket{\uparrow,1} &\rightarrow \ket*{\Psi_{\uparrow,1}} \equiv c_{11}\ket{\uparrow,1} + c_{21}\ket{\downarrow,2}, \\
	\ket{\downarrow,1} &\rightarrow \ket*{\Psi_{\downarrow,1}} \equiv c_{12}\ket{\uparrow,1} + c_{22}\ket{\downarrow,2},
\end{aligned}
\end{equation}
where $c_{ij}$ are the entries of the unitary matrix representing ${\cal C}$. By projecting onto $\ket\gamma\equiv \gamma_\uparrow\ket\uparrow+\gamma_\downarrow\ket\downarrow~(\gamma_{\uparrow,\downarrow}\in\mathbb{C})$ and imposing $\braket*{\Psi_{\uparrow,1}}{\Psi_{\downarrow,1}}=0$, we get
\begin{equation}
	 |\gamma_\uparrow|^2 c_{11}^* c_{12} +
	 |\gamma_\downarrow|^2 c_{21}^* c_{22} = 0,
\end{equation}
which is satisfied for
$\ket\gamma = (\ket\uparrow + e^{i\phi}\ket\downarrow)/\sqrt2$ for any $\phi\in\RR$.
The corresponding projection probabilities are both equal to $1/2$, as follows from
\begin{equation}
\begin{aligned}
	2\abs{\braket{\gamma}{\Psi_{\uparrow,1}}}^2 &= |c_{11}|^2 + |c_{21}|^2 = 1, \\
	2\abs{\braket{\gamma}{\Psi_{\downarrow,1}}}^2 &= |c_{12}|^2 + |c_{22}|^2 = 1.
\end{aligned}
\end{equation}
We conclude that $\on{TC}_1$ is always achievable for this class of states.
Remarkably, the freedom in the choice of the phase $\phi$ means that projections onto $\ket\pm= (\ket\uparrow \pm \ket\downarrow)/\sqrt2$ (as well as any other orthonormal basis of balanced states) are suitable to achieve entanglement transfer.
This results in an overall transfer success probability of $1$: measuring in the $\ket\pm$ basis, both of the possible outcomes achieve $\on{TC}_1$, albeit with different post-projection states.

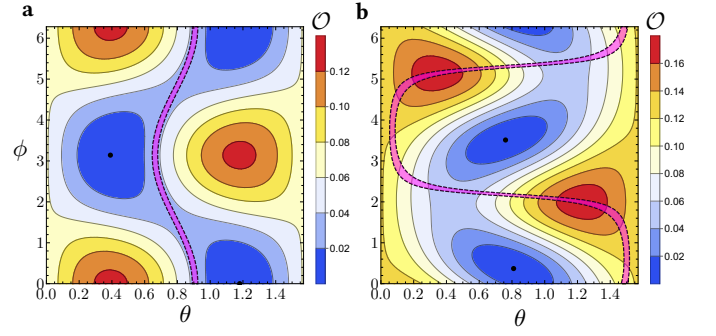
\begin{figure}[tb]
    \centering
    \subfloat[]{\begin{tikzpicture}
		\node[anchor=south west] (A) at (0, 0)%
			{\includegraphics[width=0.49\linewidth]{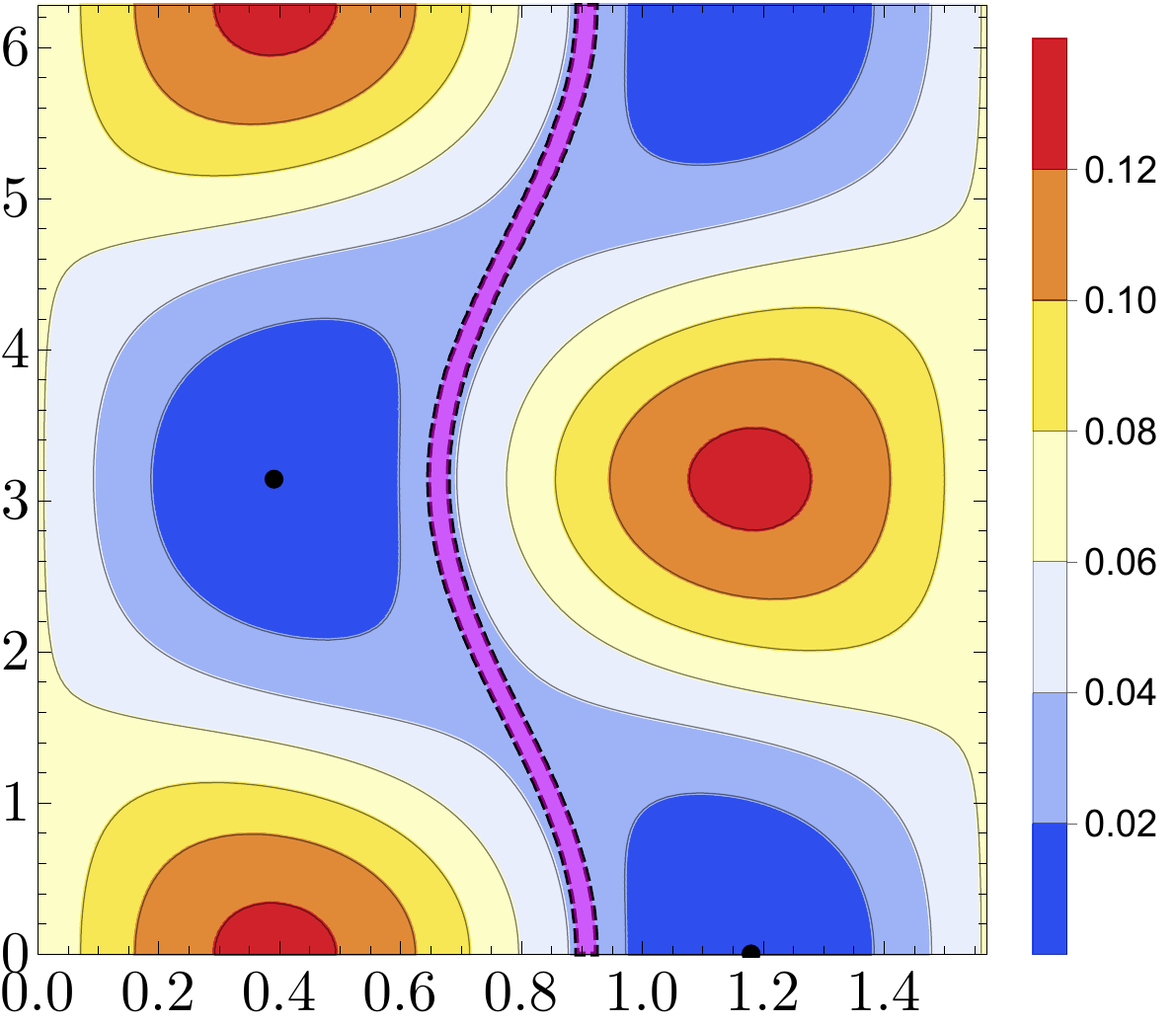}};
		\node[above, overlay] at (2.1, -.25) {$\theta$};
		\node[above, overlay] at (-0.1, 1.9) {$\phi$};
		\node[above, overlay] at (3.9, 3.65) {$\mathcal O$};
		\node[overlay] at (0, 4) {\textbf{a}};
	\end{tikzpicture}}  
    \subfloat[]{\begin{tikzpicture}
		\node[anchor=south west] (A) at (0, 0)%
			{\includegraphics[width=0.49\linewidth]{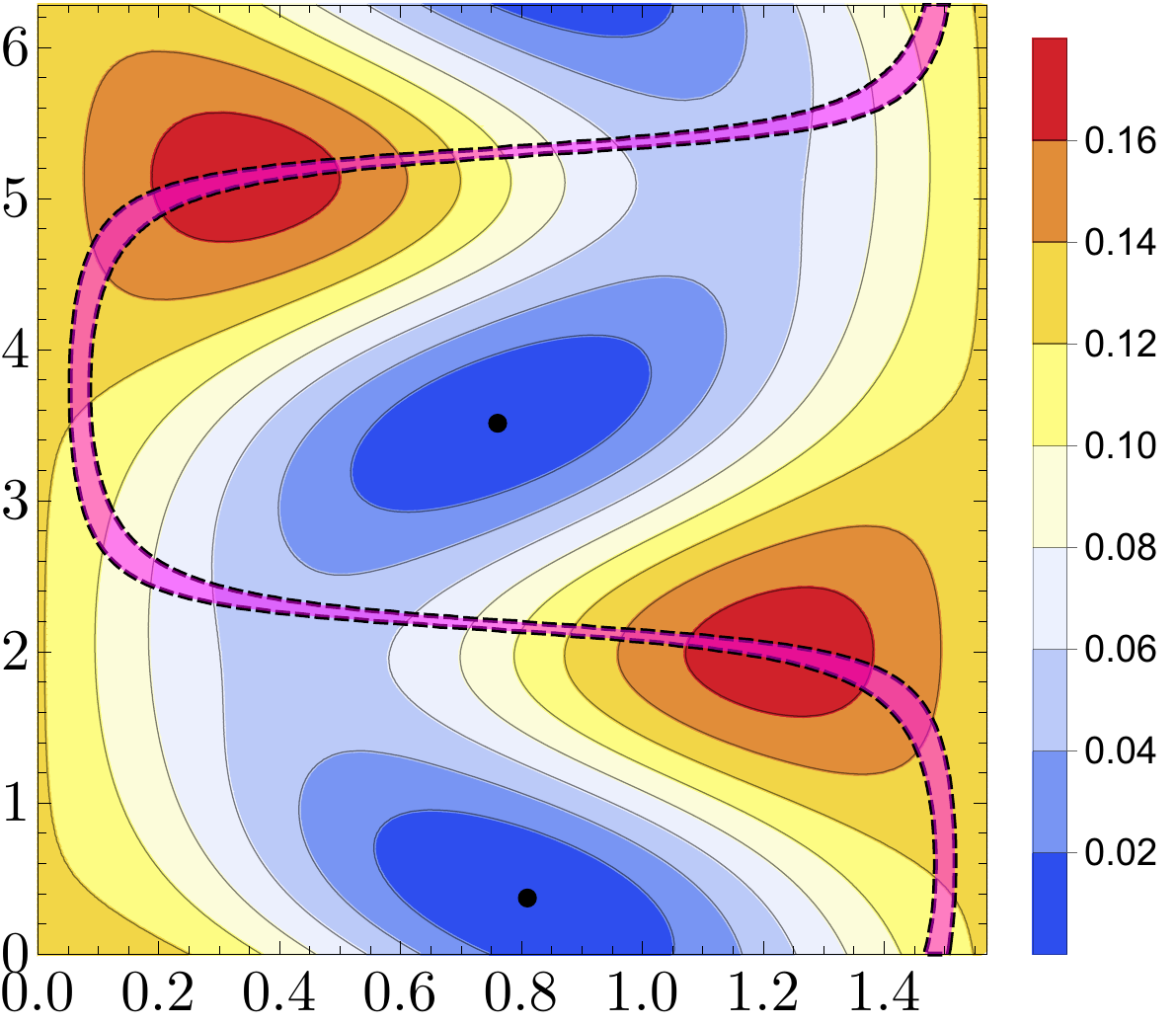}};
		\node[above, overlay] at (2.1, -.3) {$\theta$};
		\node[above, overlay] at (3.9, 3.65) {$\mathcal O$};
		\node[overlay] at (0, 4) {\textbf{b}};
	\end{tikzpicture}}
    \caption{
    Final overlap and projection probabilities for the different possible projections $\ket\gamma=\cos(\theta)\ket0+\sin(\theta)e^{i\phi}\ket1$, computed on the output of a Hadamard \textbf{(a)} or random QW \textbf{(b)} with $4$ steps. For the random QW, a random coin is used at each step.
    In each case, we consider the input states $\ket{\uparrow,1}$ and $\ket{\downarrow,1}$, and verify the satisfiability of $\on{TC}_1$ on the corresponding outputs.
    We first plot the squared overlap $\mathcal O\equiv \|\braket{\Psi_\uparrow}{\gamma}\braket{\gamma}{\Psi_\downarrow}\|^2$ for all possible $\ket\gamma$, where $\ket{\Psi_\uparrow},\ket{\Psi_\downarrow}$ denote the output states.
    We find that there are two orthogonal projections $\ket{\gamma_1},\ket{\gamma_2}$ such that this quantity is zero, represented in the figure with black dots.
    As discussed in~\cref{sec:entanglement_transfer_local_projections}, the vanishing overlap is only a necessary, not sufficient condition. To achieve $\on{TC}_1$, we also require the projection probabilities being equal, \textit{i.e.} $p_\uparrow=p_\downarrow$ where $p_s=\|\braket{\Psi_s}{\gamma}\|^2$, $s\in\{\uparrow,\downarrow\}$.
    We represent $(\theta,\phi)$ for which this condition is satisfied with the magenta region bounded by dashed black lines.
    More precisely, the magenta region outlines the set of $(\theta,\phi)$ such that the entropy of the projections probabilities, $S((p_\uparrow, p_\downarrow))$, is larger than $0.693$ (remembering that $-\ln2\simeq 0.6931$).
    It is worth noting that, while it is not in general true that $p_\uparrow+p_\downarrow=1$ for an arbitrary unitary evolution, this is always the case for QWs, which allows us to quantify how close $p_\uparrow$ and $p_\downarrow$ via the corresponding entropy.
    As clear from the figure, in these two cases, $\on{TC}_1$ cannot be achieved for any $\ket\gamma$, as the two necessary conditions cannot be simultaneously satisfied.
    }
    \label{fig:contourPlot_overlapAndEntropies}
\end{figure}

\begin{figure}[tb]
    \centering
    \includegraphics[width=\columnwidth]{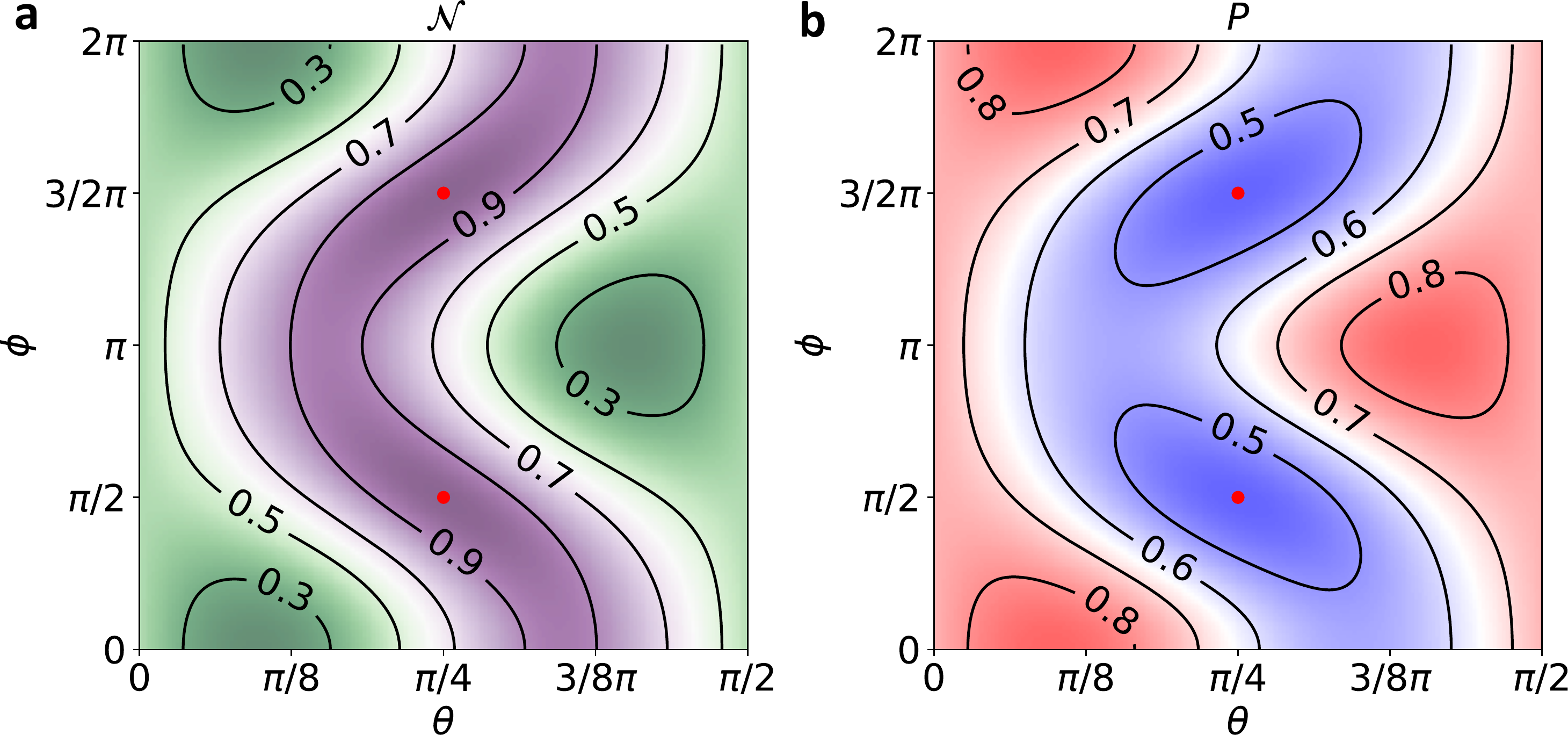}
    \caption{
        \textbf{(a)} Log-negativity $\mathcal{N} $ of the output state after 4-step Hadamard QW and coin projections along $\ket{\gamma}=\cos{\theta}\ket{\uparrow}+ e^{i \phi} \sin{\theta}\ket{\downarrow}$, with $\theta \in [0, \pi/2]$ and $\phi \in [0, 2\pi]$. States with maximum entanglement $\mathcal{N}=1$ are found for values of $\theta$ and $\phi$ identified by the red spots in the figure. In this scenario, such states are generated with probability $p$, reported in  panel \textbf{(b)} The probability of transfer is $p = 0.43$ (see the red spots highlighted in the map). This estimate takes into account that the projections on $\ket{\gamma^{\perp}}$ produce states with the same log-negativity.
    }
    \label{fig:10steps_doubleprojections_results}
\end{figure}

\parTitle{Multiple steps}
Consider now the state after multiple QW steps. 
The final reduced state on $\calH^{(1)}$ is a mixture of $\ket{\Psi_\uparrow}$ and $\ket{\Psi_\downarrow}$, where
\begin{equation}
\begin{aligned}
	\ket{\Psi_s} =
	\cos(\theta_s) \ket{\uparrow,\Psi_{s,\uparrow}} +
	\sin(\theta_s) \ket{\downarrow,\Psi_{s,\downarrow}},
\end{aligned}
\end{equation}
with $\theta_s$ and $\ket{\Psi_{s,p}}$ depending on the number of steps and choice of coin operators, and $s,p\in\{\uparrow,\downarrow\}$.
To assess the achievability of $\text{TC}_1$ we consider, as in~\cref{subsec:generalsolution_firstiteration_2Dcoin}, the matrix
$M\equiv \tr_{\calW}(\ketbra{\Psi_\uparrow}{\Psi_\downarrow})$.
This has the form
\begin{equation}\scalebox{0.97}{$\displaystyle
	M = \begin{pmatrix}
		\cos(\theta_\uparrow)\cos(\theta_\downarrow) \calO_{\uparrow\uparrow} &
		\cos(\theta_\uparrow)\sin(\theta_{\downarrow}) \calO_{\downarrow\uparrow} \\
		\cos(\theta_\downarrow)\sin(\theta_{\uparrow}) \calO_{\uparrow\downarrow} &
		\sin(\theta_\uparrow)\sin(\theta_\downarrow) \calO_{\downarrow\downarrow}
	\end{pmatrix}.
$}\end{equation}
with
$\calO_{sp}\equiv\braket{\Psi_{\downarrow s}}{\Psi_{\uparrow p}}$.
Such $M$ is not in general Hermitian, nor normal. Consequently, while it is always possible to find a state $\ket\gamma$ upon which to perform a projection, the corresponding projection probabilities are not in general equal, as shown in Fig.~\ref{fig:contourPlot_overlapAndEntropies}.
{It is worth stressing that this does not imply the impossibility of accumulating entanglement using these types of QWs. Rather, this result tells us that this is only possible via \textit{entanglement distillation}, and thus there cannot be a \textit{deterministic} protocol achieving such entanglement transfer.
In other words, Fig.~\ref{fig:contourPlot_overlapAndEntropies} shows that, in such cases, there is no projection preserving entanglement in the residual space $\HW^{(1)}\otimes\calH^{(2)}$. Nonetheless, there might still be a $\ket\delta\in\HC^{(2)}$ such that the second projection recovers the original amount of entanglement, but this can only be done probabilistically, as shown in Refs.~\cite{nielsen2001majorization,vidal1999entanglement}.
To further highlight this point, we provide in Fig.~\ref{fig:10steps_doubleprojections_results} numerical results regarding the possibility of probabilistic entanglement transfer when both projections are considered. In these cases, probabilistic entanglement transfer is possible despite $\text{TC}_1$ and $\text{TC}_2$ are not satisfied.}

There are nonetheless QW dynamics in which $\on{TC}_1$ is achievable. For example, consider a QW in which the coin is always taken to be identity: ${\cal C}=I$ at all steps. Then, after $n$ steps, the evolution amounts to
\begin{equation}
    \ket{\uparrow,1} \to \ket{\uparrow, 1}, \quad
    \ket{\downarrow,1} \to \ket{\downarrow, n},
\end{equation}
where $\ket k$ denotes the $k$-th walker position.
The matrix $M$ is thus in this case easily seen to be $M=0$, implying that the orthogonality requirement is always satisfied, making $\on{TC}_1$ achievable as long as the projection probabilities are equal. This constraint is satisfied by any balanced projection of the form $\ket\gamma=(\ket{\uparrow}+e^{i\phi}\ket{\downarrow})/\sqrt2$, $\phi\in\RR$.

\section{Entanglement accumulation}
\label{sec:entanglement_accumulation}

\parTitle{Section overview}
Here we investigate whether the entanglement transfer procedure can be applied iteratively, accumulating more and more entanglement into the state of the walkers' degrees of freedom.
For this purpose, after each successful entanglement transfer stage, which produces a state of the form
\begin{equation}
	(\ket{\gamma_1}\otimes\ket{\gamma_2})_{\calC}\otimes \ket{\Psi}_{\calW},
\end{equation}
we apply an operation restoring the entanglement between the coins, thus producing a state of the form
	$\ket{\Psi}_{\calW}\otimes\ket{\Phi}_{\calC}$,
with $\ket{\Phi}\in\HC^{(1)}\otimes\HC^{(2)}$ some entangled state --- usually a maximally entangled one.
The QW evolution is then used to correlate each coin and walker degree of freedom locally, in order to make transfering the entanglement via local projections possible.

\parTitle{Output of single round of entanglement transfer}
Suppose one round of entanglement transfer was executed successfully.
We therefore have entanglement in the bipartition $\HW^{(1)}\otimes\HW^{(2)}$, while the coin spaces are separated.
Can we perform another round of QW evolutions to transfer even more entanglement to the walkers?

\parTitle{Entanglement accumulation, problem statement}
Let us consider, as an example, the case where $\ket{\Psi}_\calW$ has entanglement dimension $2$, and the full state has the form
\begin{equation}
	 \ket{\Psi} =
	\ket{\uparrow\uparrow}_{\calC}
	\otimes
	(\ket{\psi_1\psi_2} + \ket{\psi_2\psi_1})_{\calW}/\sqrt2,
\end{equation}
for some walker states $\ket{\psi_i}$ with $\braket{\psi_i}{\psi_j}=\delta_{ij}$.
Restoring the entanglement between the coins we get
\begin{equation}
	\ket{\Psi'} =
	(\ket{\uparrow\uparrow}+ \ket{\downarrow\downarrow})_{\calC}
	\otimes
	(\ket{\psi_1\psi_2} + \ket{\psi_2\psi_1})_{\calW}/2.
	\label{eq:ent_acc_initial_state}
\end{equation}
Let us, as in~\cref{sec:entanglement_transfer_local_projections}, focus on the transferability in $\calH^{(1)}$.
The reduced state $\rho^{(1)}\equiv\tr_2\PP_{\Psi'}$ has the form
\begin{equation}
	\rho^{(1)} =
	(\PP_\uparrow + \PP_\downarrow)_\calC
	\otimes
	(\PP_{\psi_1} + \PP_{\psi_2})_\calW/4.
\end{equation}
A QW evolution $\calW_{\calS}$ then gives
$\PP[\calW_{\calS}\ket{\uparrow,\psi_1}] + 
	\PP[\calW_{\calS}\ket{\uparrow,\psi_2}] + 
	\PP[\calW_{\calS}\ket{\downarrow,\psi_1}] + 
	\PP[\calW_{\calS}\ket{\downarrow,\psi_2}]= \PP_{\Psi_1} + \PP_{\Psi_2} + \PP_{\Psi_3} + \PP_{\Psi_4}$,
where $\braket{\Psi_i}{\Psi_j}=\delta_{ij}$, and thus $\calW_{\calS}\rho^{(1)}\calW_{\calS}^\dagger$ has rank $4$.
Achieving entanglement transfer now entails finding $\ket\gamma\in\HC^{(1)}$ such that
\begin{equation}
	\mel{\Psi_i}{\PP_\gamma\otimes I_\calW}{\Psi_j} = \delta_{ij} p_{\on{proj}}.
\end{equation}
Each successive entanglement transfer iteration involves a doubling of the number of orthogonal states to preserve, as follows from observing that if $A$ has rank $r$ and $B$ has rank $r'$, then $A\otimes B$ has rank $rr'$.

\parTitle{QWs with $C=I$}
Consider now a QW in which each coin operation is the identity: $\calC=I$.
We will show that, with this particular type of dynamics, we can accumulate arbitrary amounts of entanglement into the walkers' degrees of freedom, using the coins as mediators.
The unitary evolution corresponding to $n$ steps with $\calC=I$ is $\calW_{\calS,n}=\calS^n$ with $\calS$ the controlled-shift operation.
The action on the basis states is then
\begin{equation}
    \calS^n = \PP_\uparrow \otimes I + \PP_\downarrow\otimes \EE_+^n,
\end{equation}
where $\EE_+\equiv \sum_k\ketbra{k+1}{k}$ is the operation shifting the walker's position, and $\EE_+^n$ is thus the operator moving the walker $n$ positions forward.
Consider an initial state
\begin{equation}
    (\ket{\uparrow,\uparrow}+\ket{\downarrow,\downarrow})_\calC\otimes
    \ket{\Psi}_\calW
\end{equation}
with $\ket{\Psi}_\calW$ an entangled state of the walkers in which the difference between final and initial occupied positions is $\ell\in\NN$
(for example, if $\sqrt2\ket{\Psi}_\calW=\ket{1,1}+\ket{3,3}$, then $\ell=2$).
If $\ket{\Psi}_\calW$ has rank $r$, the reduced state on $\calH^{(1)}$ has the form
\begin{equation}
    \tr_2(\PP_{\Psi}) = \frac12 (\PP_\uparrow+\PP_\downarrow)\otimes\sum_{k=1}^r p_k \PP_{\psi_k}
\end{equation}
for some set of orthonormal states $\{\ket{\psi_k}\}_k\subset\HW^{(1)}$.
Evolving through $\calS^{\ell+1}$, we get
\begin{equation}
    \tr_2(\PP_{\Psi}) \to
    \frac12\left(
        \PP_\uparrow\otimes\sum_{k=1}^r p_k \PP_{\psi_k} +
        \PP_\downarrow\otimes\sum_{k=1}^r p_k \PP_{\psi_k'}
    \right),
\end{equation}
with $\braket{\psi_j'}{\psi_k'}=\delta_{jk}$ and $\braket{\psi_j'}{\psi_k}=0$.
Then, any balanced projection $\ket\gamma=\frac{1}{\sqrt2}(\ket \uparrow+e^{i\phi}\ket \downarrow)$ achieves $\on{TC}_1$, which means that the entanglement can be transferred deterministically from coins to walkers. 

\begin{figure}[t]
    \centering
    \includegraphics[width=\columnwidth]{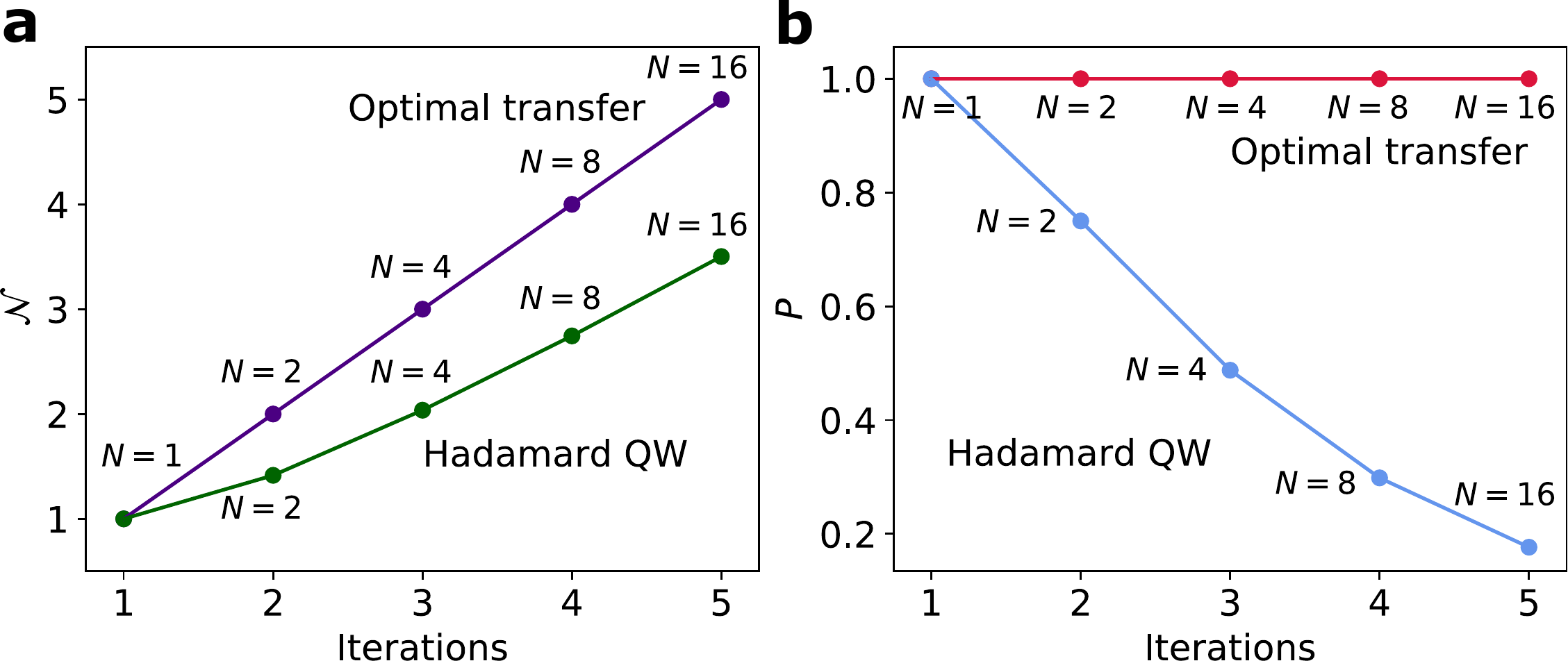}
    \caption{
        \textbf{(a)} Trends of $\mathcal{N}$ for QWs with $\mathcal{C}=\mathcal{I}$ (purple) and Hadamard QWs (green) in the entanglement accumulation protocol. The numbers near the markers specify the number of QW steps needed to store the entanglement in the walkers subspaces. 
        The first case corresponds to the deterministic optimal transfer described in the main text, in which one ebit is transferred at each iteration. In the Hadamard QWs this optimal transfer it is not achievable after the first iteration. \textbf{(b)} Accumulation probability for the two cases.
    }
    \label{fig:fig_accumulation}
\end{figure}

\parTitle{Specific optimal accumulation protocol}
In light of these findings, we can now propose the following explicit protocol, which allows to accumulate deterministically entanglement into the walkers degrees of freedom using the coins as mediators.
Starting from the state
$(\ket{\uparrow,\uparrow}+\ket{\downarrow,\downarrow})/\sqrt2\otimes\ket{1,1}\in\calH$,
we apply the conditional shift operation to both QWs and then measure both coins in the basis $\ket\pm$. 
The possible states after the projection are then
   $(\ket{1,1} \pm \ket{2,2})/\sqrt2$,
where the sign is $+$ if the two coins are found in the same state, and $-$ otherwise.
Restoring the entanglement between the coins, we then re-apply the QW evolution, now for two steps, and project again in the basis $\ket\pm$, resulting in an output state of the form
\begin{equation}
    \frac12[(\ket{1,1}\pm\ket{2,2}) \pm (\ket{3,3} \pm \ket{4,4})].
\end{equation}
This procedure can be iterated to accumulate more and more entanglement in $\HW^{(1)}\otimes\HW^{(2)}$.
At the $n$-th iteration, we evolve both 
systems through $2^n$ QW steps with $\calC=I$, that is, through the unitary $\calS^{2^n}\otimes\calS^{2^n}$, and then project onto the $\ket\pm$ basis,
resulting in a maximally entangled state of the form
\begin{equation}
    2^{-n/2}\sum_{k=1}^{2^n} (-)^{\sigma_k}\ket{k,k},
\end{equation}
with $(-)^{\sigma_k}\in\{1,-1\}$ for all $k$.
In Fig.\ref{fig:fig_accumulation} we report the trend of the deterministic transfer and accumulation described above. Notice that this goal cannot always be achieved, as for example in the case of the Hadamard QW reported in the figure. 
Here it is not possible to transfer one-ebit of entanglement per iteration, not even probabilistically.

\parTitle{Numerical results \label{transfer}}

\section{Entanglement retrieval}
\label{sec:entanglement_retrieval}

The arguments of~\cref{sec:entanglement_transfer_local_projections} do not make assumptions on the dimensions of $\HC^{(i)}$ and $\HW^{(i)}$. This means that they can be used not only to study the transfer of entanglement from coins to positions, but also the other way around.
For example, if the initial reduced state on $\calH^{(1)}$ is
    $[\PP_\uparrow\otimes (\PP_1+\PP_2)]/2$,
with $\ket1,\ket2$ a pair of orthonormal walker's states, then applying a Hadamard operation to the coin, and two steps of QW evolution with $\calC=I$, we obtain the state
\begin{equation}
    \frac14[\PP_\uparrow\otimes(\PP_1+\PP_2) + \PP_\downarrow\otimes(\PP_3+\PP_4)].
\end{equation}
Then, measuring in the Hadamard four-dimensional basis --- \textit{i.e.}, the basis formed by the columns of the $4\times4$ Hadamard matrix --- we achieve $\on{TC}_1$.

\section{Experimental proposal}
\label{sec:experimental_proposal}
QWs have been previously demonstrated on photonic platforms~\cite{Perets2008, Peruzzo2010, broome2010bdp, Rohde_2011, Schreiber2010,  sansoni2012quantum, boutari2016time, cardano2015quantum, cardano2016statistical, caruso2016maze}.
We propose here a scheme to implement our entanglement transfer and accumulation protocol in an optical platform, encoding coin and walker degrees of freedom in circular polarization and OAM degrees of freedom of single photons.

Implementing the entanglement transfer protocol involves as starting point two polarization-entangled photons, that can be generated via photon sources based on parametric down conversion. Then, each photon of the pair evolves through a QW evolution in the polarization and OAM degrees of freedom, followed by a projective measurement on the polarizations.
The coin operators are realized on the polarization through suitable sets of waveplates. The shift operator, which involves an interaction between OAM and polarization, is naturally implemented by the inhomogeneous and birefringent devices known as q-plates \cite{marrucci2006optical,giordani2019experimental}.
Projective measurements on the coins are realized using waveplates and a polarizing beam splitter.

For the entanglement accumulation protocol, one also needs a way to ``\textit{reload}'' the entanglement into the photons' polarization without affecting their OAMs.
As discussed in~\cref{sec:entanglement_accumulation}, the first entanglement transfer procedure results in one of the states
\begin{equation}
\label{eq:26}
    \ket{++}\otimes(\ket{1,1}\pm\ket{2,2})/\sqrt2,
\end{equation}. 
where $\vert 1 \rangle$ and $\vert 2 \rangle$ label OAM states, while $\vert + \rangle$ are diagonal polarization states. It is straightforward to show that the action of a polarizing beam-splitter combined with two half-waveplates can restore, with probability $1/2$, the entangled state in polarization needed to achieve accumulation.
Expressing the state of Eq. \eqref{eq:26} in terms of creation operators $a^{\dagger}$ and $b^{\dagger}$ of the two photons, we have:
\begin{equation}
   \frac{1}{\sqrt{2}}\left(a^{\dagger}_{+, 1}b^{\dagger}_{+,1} \pm a^{\dagger}_{+, 2}b^{\dagger}_{+, 2} \right)\ket{\mathrm{vac}}, 
   \label{eq: snd_quant}
\end{equation}
where $\ket{\mathrm{vac}}$ is the vacuum state in the Fock representation. The two photons are injected in the input ports, labelled by $\{a, b\}$, of a polarizing beam-splitter, after a polarization rotation made by two half-waveplates of angles $\theta_a$ and $\theta_b$ respectively. The creation operators after the overall transformation become
\begin{equation}
\begin{aligned}
    a^{\dagger}_{+, 1/2} & \rightarrow \cos{\theta_a}a^{\dagger}_{+, 1/2} + \imath \sin{\theta_a}b^{\dagger }_{-, 1/2} \\
   b^{\dagger}_{+, 1/2} & \rightarrow \cos{\theta_b}b^{\dagger}_{+, 1/2} + \imath \sin{\theta_b}a^{\dagger}_{-, 1/2}
\end{aligned}\end{equation}

Substituting such expression in Eq. \eqref{eq: snd_quant} and choosing the orientation of the two half-waveplates $\theta_a = \theta_b= \pi/4$ we obtain that the output state is composed by two terms. The first term corresponds to the two photons exiting from different output ports of the polarizing beam-splitter, while the second term corresponds to the case where the photons exit from the same port.
The first part of this state embodies the resource needed for the protocol accumulation, and has the following form:
\begin{align}
\frac{\left( \ket{+ +} \pm
\ket{- -}\right) }{\sqrt{2}}\otimes \frac{\left( \ket{1,1} \mp
\ket{2,2}\right) }{\sqrt{2}}.
\end{align}
We can discard the second term where two photon exit from the same port by post-selecting two-fold coincidences between single-photon detectors at the end of the second iteration. It is worth noting that the probabilistic generation of the second maximally entangled state is due to the choice of encoding qubits in photons.  However, we remark that we could also consider the state produced by the projection $\ket{- -}$ after the first operation. Indeed, this projection produces
states with the same symmetry properties. In this way it is possible to double the probability of generating states with more than one-ebit of entanglement.

{As a final remark, we note that the whole protocol, being essentially based upon the general QW dynamics, can be implemented in various experimental platforms~\cite{Manouchehri2014,cote2006quantum,Schmitz2009,Zahringer2010,Meinert2014}. }

\section{Conclusions}

We have addressed the generation of high-dimensional entangled states through a protocol of entanglement transfer from a low-dimensional resources. We have identified general transfer conditions that, if met, guarantee the successful pouring of any entanglement initially contained in the state of the resource to the high-dimensional {\it receiver}. This has then allowed us to draw a specific analysis aimed at the dynamics entailed by a QW, where low-dimensional resources and high-dimensional receivers are naturally embodied by coin and walker degrees of freedom respectively. While characterizing the performance of the entanglement transfer scheme, we have been able to design schemes for entanglement accumulation and retrieval, thus drawing a complete picture for the manipulation of entanglement through a hetero-dimensional interface of great experimental potential. Indeed, the QW-based protocols addressed and studies in this paper are fully amenable to an implementation making use of polarization and OAM encoding. The scenario set by our schemes sets a promising framework for the use of low-dimensional entanglement as a resource to achieve otherwise complex entangled structures and states that can be experimentally synthesised and exploited.

\begin{acknowledgments}

This work is supported by MIUR via PRIN 2017 (Progetto di Ricerca di Interesse Nazionale): project QUSHIP (2017SRNBRK), 387439, and by the ERC Advanced Grant QU-BOSS (Grant agreement no. 884676). The authors acknowledge financial support from H2020 through the Collaborative Project TEQ (Grant Agreement No.  766900), the DfE-SFI Investigator Programme (Grant No. 15/IA/2864), the Leverhulme Trust Research Project Grant UltraQute (grant nr.~RGP-2018-266), COST Action CA15220, and the Royal Society Wolfson Research Fellowship scheme (RSWF\textbackslash R3\textbackslash183013). T.G. acknowledges La Sapienza University of Rome via the grant for joint research projects for the mobility n.2289/2018 Prot. n.50074
\end{acknowledgments}

\appendix

\section{Entanglement decreases if orthogonality is not preserved}
\label{appA}
Let us show that if $\braket{\tilde u_j}{\tilde u_k}\neq\delta_{jk}$ then the Schmidt coefficients must change upon projection.
Indeed, in this case, $\ket{\Psi_\gamma}$ has the form
$\ket{\Psi_\gamma}=\sum_k \sqrt{\tilde p_k} \ket{\tilde u_k}\ket{v_k}$ where $\braket{v_j}{v_k}=\delta_{jk}$ and $\sum_k \tilde p_k=1$.
Denoting with $\Psi_\gamma$ the matrix whose vectorization is $\ket{\Psi_\gamma}$, this Schmidt decomposition amounts to the singular value decomposition $\Psi_\gamma = U\sqrt DV^\dagger$, with $D=\on{diag}(\tilde p_1,...,\tilde p_n)$, $V$ the unitary matrix whose columns are $\ket{v_k}$, and $U$  the (non-unitary) matrix with columns $\ket{\tilde u_k}$.
Then
\begin{equation}
    \Psi_\gamma\Psi_\gamma^\dagger = U D U^\dagger
    = \sum_k \tilde p_k \PP_{\tilde u_k},
\end{equation}
where $\PP_{\tilde u_k}=\ketbra{\tilde u_k}$ are in general non-orthogonal rank-$1$ projectors.
Let us then prove that if a matrix is a convex combination of rank-$1$ projections, then it always majorizes the vector of coefficients of the convex combination.
In our case, this translates to $\sum_k \tilde p_k\PP_{\tilde u_k}\succeq \tilde{\bs{p}}$.

Let $P_k$ be rank-$1$ projections, $p_k\ge0$ coefficients such that $\sum_{k=1}^n p_k=1$, and
$A\equiv \sum_{k=1}^n p_k P_k$. We want to prove that $A\succeq \bs p$,
where $\bs p=(p_k)_{k=1}^n$ is the vector of coefficients, and the majorization relation is defined on Hermitian matrices via the corresponding vector of eigenvalues, that is,
$A\succeq\bs p\Longleftrightarrow \bs\sigma(A)\succeq\bs p$ where $\bs\sigma(A)$ is the vector of eigenvalues of $A$. If $A$ has dimension larger than $n$, we define $\bs\lambda(A)$ as the vector of the $n$ largest eigenvalues, in order to make the majorization relation well-defined.
Without loss of generality, let us assume that the $p_k$ are in decreasing order: $p_1 \ge p_2 \ge ...\ge p_n$.
Define the partial sums $A_\ell\equiv \sum_{k=1}^\ell p_k P_k$, so that $A=A_n$.
Observe that $A_\ell \ge A_r$ whenever $\ell\ge r$.
Because $\rank(P_k)=1$ for all $k$, we must also have $\rank(A_\ell)\le \ell$.
Denoting with $\lambda_j^\downarrow(A)$ the $j$-th largest eigenvalue of $A$, this implies that
\begin{equation}
    \sum_{k=1}^\ell \lambda_k^\downarrow(A_\ell) = \tr(A_\ell)
    = \sum_{k=1}^\ell p_k.
\end{equation}
Using $A=A_n\ge A_\ell$ for all $1\le \ell< n$, we thus conclude that
\begin{equation}
    \sum_{k=1}^\ell \lambda_k^\downarrow(A) \ge 
    \sum_{k=1}^\ell \lambda_k^\downarrow(A_\ell)
    = \sum_{k=1}^\ell p_k \equiv \sum_{k=1}^\ell p_k^\downarrow,
\end{equation}
that is, $\bs\lambda(A)\succeq \bs p$, which is the definition of $A\succeq \bs p$.

\parTitle{Entanglement degradation for different probabilities}
Assuming $\ket{\tilde u_k}$ are orthogonal, then the Schmidt coefficients of $\ket{\Psi_\gamma}$ are $\sqrt{\tilde p_k}=p^{-1/2}_{\on{proj}}\sqrt{p_k q_k}$.

\section{Finding projections preserving orthogonality}
\label{appB}

We prove in this section that, for any state of the form $\ket\Psi=\sqrt{p_1}\ket{u,u'}+\sqrt{p_2}\ket{v,v'}$, with $\braket{u}{v}=\braket{u'}{v'}=0$, there is some $\ket\gamma$ such that the post-projected states are orthogonal, \emph{i.e.} such that $\braket*{\tilde u}{\tilde v}=0$ where $\sqrt{p_u}\ket*{\tilde u}= \braket{\gamma}{u}$ and $\sqrt{p_v}\ket*{\tilde v} = \braket{\gamma}{v}$.

Here, $\ket\Psi\in\calH^{(1)}\otimes\calH^{(2)}$, $\ket u,\ket v\in\calH^{(1)}$, and $\ket{u'},\ket{v'}\in\calH^{(2)}$.
Moreover, $\calH^{(1)}=\HC^{(1)}\otimes\HW^{(1)}$, and $\ket\gamma\in\HC^{(1)}$.
Note that here we assume $\dim(\HC^{(1)})=2$, while the only requirement on $\HW^{(1)}$ and $\calH^{(2)}$ is that their dimension must be larger than $2$, in order to accommodate $\ket\Psi$.

Define $M\equiv \tr_{\calW}(\ketbra{u}{v})\in\mathrm{Lin}(\HC^{(1)})$. Note that this is a $2\times2$ traceless matrix, as follows from $\braket{u}{v}=0$. Our objective is then to find $\ket\gamma$ such that $\mel{\gamma}{M}{\gamma}=0$.
For the purpose, we consider different scenarios:
\begin{enumerate}
	\item If $M$ is normal, then
	\begin{equation}
		M = \lambda( \ketbra{v_1} - \ketbra{v_2}),
	\end{equation}
	for some $\lambda\in\CC$ and $\braket{v_i}{v_j}=\delta_{ij}$.
	Then,
	\begin{equation}
		\sqrt2\ket{\gamma_\phi} = \ket{v_1} + e^{i\phi}\ket{v_2}, \quad \phi\in\RR
	\end{equation}
	are all suitable projections such that $\mel{\gamma_\phi}{M}{\gamma_\phi}=0$.
	Note that this also implies that we can find \emph{orthogonal} states that both correspond to valid projections.
	\item Consider now a generic $2\times2$ $M$. 
	Given a two-dimensional $M$ with $\tr(M)=0$, provided $M\neq0$, we must always have $M^2=-\det(M) I$.
	This follows from observing that the eigenvalues of $M$ are $\pm\sqrt{-\det M}$, and therefore $(M+\sqrt{-\det M})(M-\sqrt{-\det M})=0$
	Writing its singular value decomposition as
	$M=UDV^\dagger$, this implies that
	$
		UDV^\dagger UDV^\dagger = -\det(M) I,
	$ 
	and therefore
	\begin{equation}
		DV^\dagger U = -\det(M) (V^\dagger U)^\dagger D^{-1}.
	\end{equation}
	If $D=d_1\PP_1+d_2\PP_2$ and
	$V^\dagger U = \ketbra{1}{w_1} + \ketbra{2}{w_2}$, then
	\begin{equation}
	\begin{gathered}
		d_1 \ketbra{1}{w_1} + d_2 \ketbra{2}{w_2} = \\
		-e^{i\phi}(d_2\ketbra{w_1}{1} + d_1\ketbra{w_2}{2}),
	\end{gathered}
	\end{equation}
	where $\det(M)=\abs{\det(M)}e^{i\phi}$ and we observed that $\abs{\det(M)}=d_1 d_2$.
	There are then two possibilities: either $d_1=d_2$, which implies $M$ is normal, and this case was covered above, or $d_1\neq d_2$, which implies by the uniqueness of the singular value decomposition that $\ket{w_1}=\ket2$ and $\ket{w_2}=\ket1$ up to phases.
	Consequently, we would have
	\begin{equation}
		M = d_1 \ketbra{u_1}{v_1} + d_2 \ketbra{u_2}{v_2},
	\end{equation}
	where $\braket{u_i}{u_j}=\braket{v_i}{v_j}=\delta_{ij}$ and $\braket{u_1}{v_2}=\braket{u_2}{v_1}=0$.
	We can then use $\ket\gamma=\ket{v_i}$ as suitable projections, as $\mel{v_i}{M}{v_i}=0$.
\end{enumerate}

\section{Entanglement transfer toy examples}

We give in this section a few toy examples showcasing the use of the results presented in~\cref{appA,appB}.
\subsection{Example with different projection probabilities}
Suppose
\begin{equation}
\begin{aligned}
    2\ket*{u} \equiv \left(
        \sqrt2\ket\uparrow \otimes\ket2 +
        \ket\downarrow \otimes(\ket1 + \ket2)
    \right), \\
    2\ket*{v} \equiv \left(
        \ket\uparrow\otimes(\ket1 + \ket2) -
        \sqrt2\ket\downarrow\otimes \ket2
    \right).
\end{aligned}
\end{equation}
Then,
$M = \frac{1}{2\sqrt2}\begin{pmatrix}1 & -\sqrt2 \\ \sqrt2 & -1 \end{pmatrix}$.
The singular values of $M$ are $\sqrt2\ket{\gamma_\pm}=\ket \uparrow\pm\ket \downarrow$, which are therefore also the projections that maintain the orthogonality.
The corresponding projection probabilities are
\begin{equation}
\begin{aligned}
	p_\pm^u
	&= \lvert\braket{\gamma_\pm}{u}\rvert^2
	= (2\pm\sqrt2)/4, \\ 
	p_\pm^v
	&= \lvert\braket{\gamma_\pm}{v}\rvert^2
	= (2\mp\sqrt2)/4.
\end{aligned}
\end{equation}
and the projected states read
\begin{equation}
\begin{aligned}
	\sqrt{p_\pm^u} \ket*{\tilde u_\pm} \equiv
		\braket{\gamma_\pm}{u}, \qquad
	\sqrt{p_\pm^v} \ket*{\tilde v_\pm} \equiv \braket{\gamma_\pm}{v}.
\end{aligned}
\end{equation}
It follows that, despite $\ket{\gamma_\pm}$ preserving the orthogonality of $\ket u,\ket v$, the two states correspond to different projection probabilities, and therefore entanglement is necessarily degraded.
More explicitly, using $\ket{\gamma_+}$ as an example, the corresponding projection probabilities are $p^u_+=(2+\sqrt2)/4$ and $p^v_+=(2-\sqrt2)/4$.
This means that if we have an entangled state of the form
\begin{equation}
	\ket\Psi = \sqrt{p_1} \ket{u}\otimes\ket0 + \sqrt{p_2}\ket v\otimes\ket1,
\end{equation}
where $\ket0,\ket1$ is an arbitrary pair of orthonormal states in an auxiliary space, then projecting onto $\ket{\gamma_+}$ gives the state
\begin{equation}
	N^{-1/2}(\sqrt{p_1 p^u_+} \ket*{\tilde u_+}\otimes\ket0 +
	\sqrt{p_2 p^v_+} \ket*{\tilde v_+}\otimes\ket1)
\end{equation}
with probability $N\equiv p_1 p_+^u + p_2 p_+^v$.
Clearly, because $p_+^u\neq p_+^v$, the Schmidt coefficients of this states are different, and thus the state is less entangled.

\subsection{Example with same projection probabilities}
Suppose
\begin{equation}
\begin{aligned}
    2\ket*{u} &=
    	\sqrt2\ket\uparrow\otimes \ket 2 +
    	\ket\downarrow\otimes (\ket1 + \ket2), \\
    2\ket*{v} &=
    	\ket\uparrow\otimes (\ket1 - \ket2) +
    	\sqrt2\ket\downarrow\otimes \ket 1.
\end{aligned}
\end{equation}
Then,
$M = \frac{1}{2\sqrt2}\begin{pmatrix}
    -1 & 0 \\
    0 & 1
\end{pmatrix}$ is normal with eigenvectors $\ket*{\lambda_+}=\ket\uparrow$, $\ket*{\lambda_-}=\ket\downarrow$.
It follows that any balanced state of the form
$\sqrt2\ket{\gamma_\phi}=\ket\uparrow + e^{i\phi}\ket\downarrow$ preserves the orthogonality of $\ket u, \ket v$.
Correspondingly, we have
\begin{equation}
\begin{aligned}
	2\sqrt2 e^{i\phi}\braket{\gamma_\phi}{u} &=
		\ket1 + (\sqrt2 e^{i\phi}+1)\ket2, \\
	2\sqrt2 e^{i\phi}\braket{\gamma_\phi}{v} &=
		(\sqrt2 + e^{i\phi})\ket1 - e^{i\phi}\ket2,
\end{aligned}
\end{equation}
with probabilities
\begin{equation}
\begin{aligned}
	p_\phi^u = p_\phi^v = \frac{1}{4}(2 + \sqrt2 \cos\phi).
\end{aligned}
\end{equation}
It follows that any $\ket{\gamma_\phi}$ achieves entanglement transfer.
Moreover, we can choose two orthogonal states, \emph{e.g.} $\ket{\gamma_0}$ and $\ket{\gamma_\pi}$, so that entanglement transfer is achieved deterministically.

\bibliography{vvb}
\bibliographystyle{apsrev4-1}
\end{document}